\documentclass[sigconf]{acmart}
\renewcommand\footnotetextcopyrightpermission[1]{} 

\usepackage{array,ragged2e}
\usepackage{amsmath,amssymb,amsfonts}
\usepackage{algorithmic}
\usepackage{graphicx}
\usepackage{textcomp}
\usepackage{xcolor}
\usepackage{booktabs}
\usepackage{multirow}
\usepackage{enumitem}  
\usepackage{verbatim}
\usepackage{subfigure}
\usepackage[font=small,skip=0pt]{caption}
\usepackage{titlesec}
\AtBeginDocument{%
  \providecommand\BibTeX{{%
    \normalfont B\kern-0.5em{\scshape i\kern-0.25em b}\kern-0.8em\TeX}}}

\acmConference{}
\acmBooktitle{}

\titlespacing*{\section}
{0pt}{.1ex}{.1ex}
\titlespacing*{\subsection}
{0pt}{.1ex}{.1ex}
\titlespacing*{\subsubsection}
{2ex}{.1ex}{1ex}

\newcolumntype{R}[1]{>{\raggedright\arraybackslash}p{#1}}

\newcommand{\zc}[1]{\textbf{\textcolor{red}{ZC:}}\textcolor{red}{#1}}

\begin{document}

\title{Task-Oriented API Usage Examples Prompting Powered By Programming Task Knowledge Graph}




\author{Jiamou Sun}
\affiliation{%
	\institution{Australia National University}
	\city{Canberra}
	\country{Australia}}
\email{u5871153@anu.edu.au}

\author{Zhenchang Xing}
\affiliation{%
	\institution{Australia National University}
	\city{Canberra}
	\country{Australia}}
\email{Zhenchang.Xing@anu.edu.au}

\author{Xin Peng}
\affiliation{%
	\institution{Fudan University}
	\city{Shanghai}
	\country{China}}
\email{pengxin@fudan.edu.cn}

\author{Xiwei Xu}
\affiliation{%
	\institution{Data61, CSIRO}
	\city{Sydney}
	\country{Australia}}
\email{Xiwei.Xu@data61.csiro.au}

\author{Liming Zhu}
\affiliation{%
	\institution{CSIRO's Data61 \& School of CSE, UNSW}
	\city{Sydney}
	\country{Australia}}
\email{Liming.Zhu@data61.csiro.au}


\begin{abstract}
	
	Programming tutorials are often created to demonstrate programming tasks with code examples.
	However, our study of Stack Overflow questions reveals the low utilization of high-quality programming tutorials, which is caused task description mismatch and code information overload.
	Document search can find relevant tutorial documents, but they often cannot find specific programming actions and code solutions relevant to the developers' task needs.
	The recently proposed activity-centric search over knowledge graph supports direct search of programming actions, but it has limitations in action coverage, natural language based task search, and coarse-grained code example recommendation.
	In this work, we enhance action coverage in knowledge graph with actions extracted from comments in code examples and more forms of activity sentences.
	To overcome the task description mismatch problem, we develop a code matching based task search method to find relevant programming actions and code examples to the code under development.
	We integrate our knowledge graph and task search method in the IDE, and develop an observe-push based tool to prompt developers with task-oriented API usage examples.
	To alleviate the code information overload problem, our tool highlights programming action and API information in the prompted tutorial task excerpts and code examples based on the underlying knowledge graph.
	Our evaluation confirms the high quality of the constructed knowledge graph, and show that our code matching based task search can recommend effective code solutions to programming issues asked on Stack Overflow.
	A small-scale user study demonstrates that our tool is useful for assisting developers in finding and using relevant programming tutorials in their programming tasks.

\end{abstract}

\maketitle

\section{Introduction}

\begin{figure*}
	\captionsetup{aboveskip=2pt}
	\centering
	\subfigure[\href{https://developer.android.com/guide/topics/ui/dialogs\#FullscreenDialog}{\textcolor{green}{Showing a Dialog Fullscreen ...}} ]{
	\captionsetup{skip=0pt}
	\begin{minipage}{0.24\linewidth}
		\vspace{-1mm}
		\includegraphics[width=\linewidth]{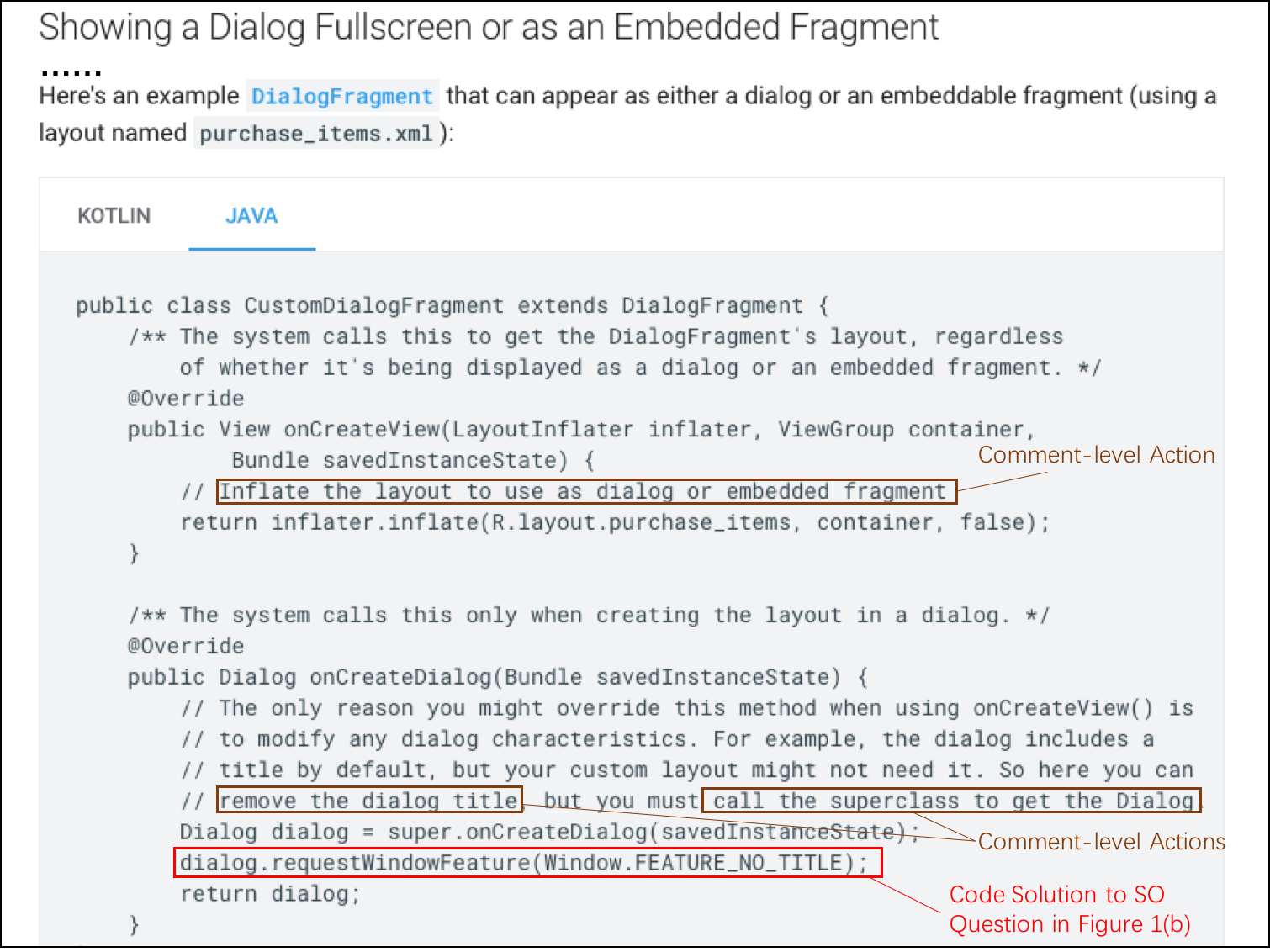}
		\vspace{-1mm}
		\label{fig:tutorialexample2}
	\end{minipage}}
	\subfigure[\href{https://stackoverflow.com/questions/2644134/android-how-to-create-a-dialog-without-a-title}{\textcolor{blue}{How to create a Dialog without a title}}]{
		\captionsetup{skip=0pt}
		\begin{minipage}{0.24\linewidth}
			\includegraphics[width=\linewidth]{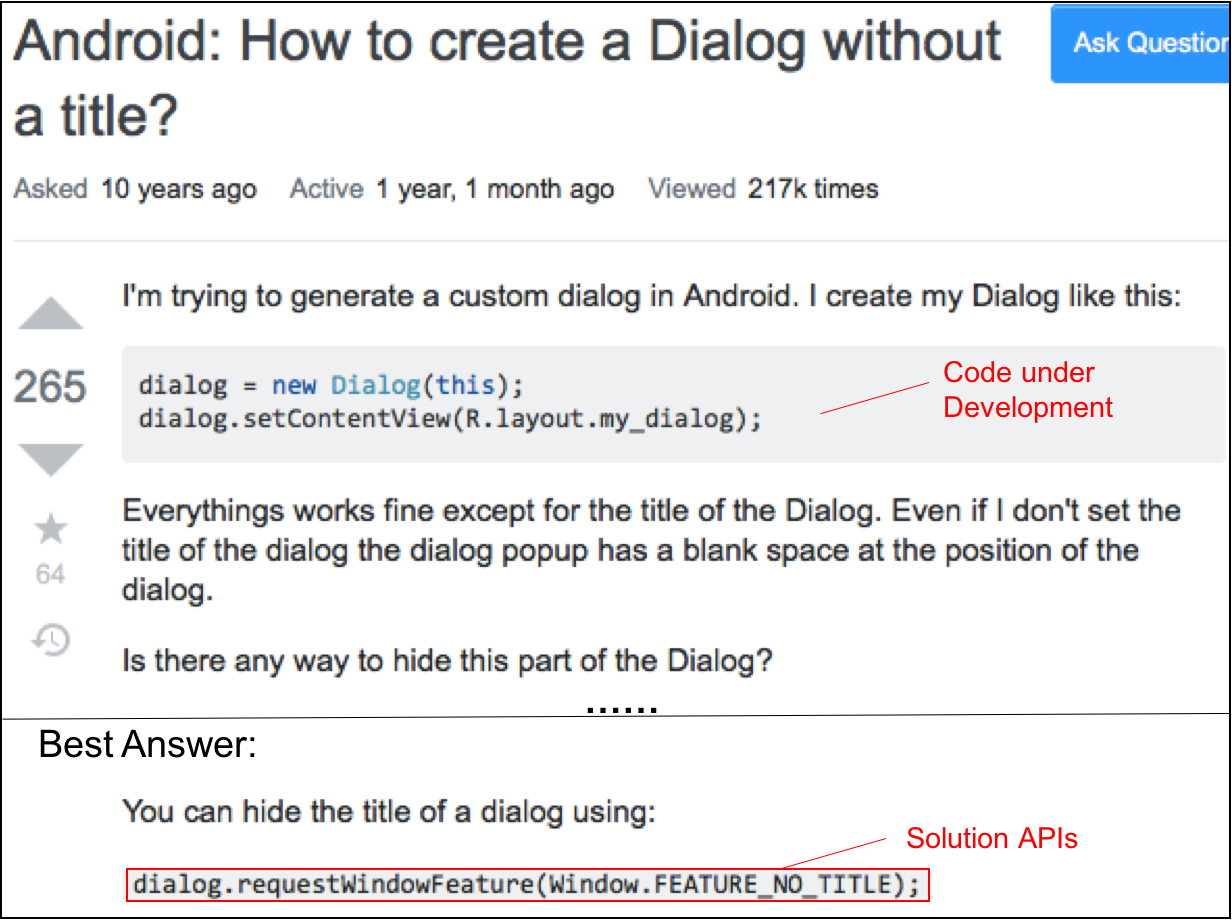}
			\vspace{-1mm}
			\label{fig:tutorialexample1}
	\end{minipage}}
	\subfigure[\href{https://developer.android.com/guide/components/fragments.html\#Transactions}{\textcolor{green}{Performing Fragment Transactions}}]{
		\captionsetup{skip=0pt}
		\begin{minipage}{0.24\linewidth}
			\vspace{-1mm}
			\includegraphics[width=\linewidth]{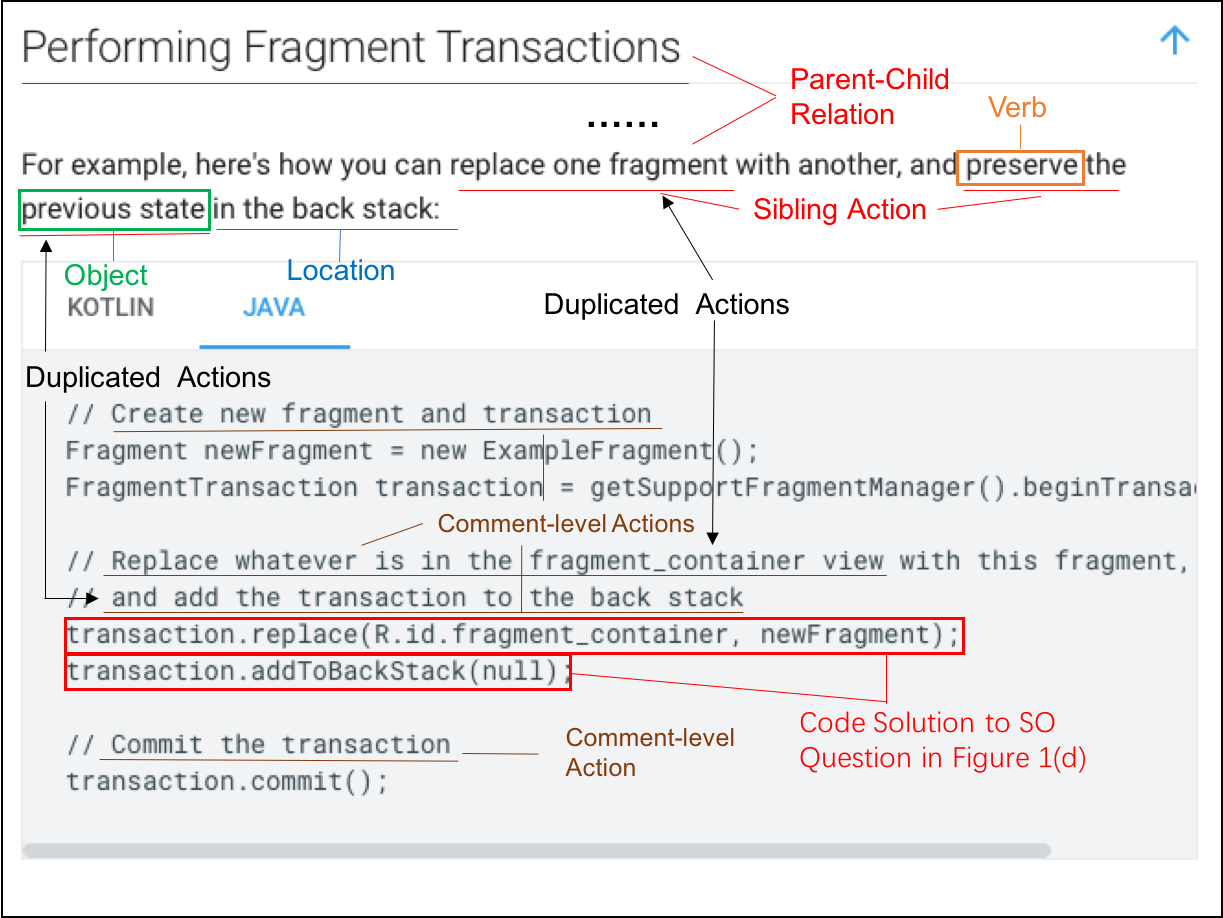}
			\vspace{-1mm}
			\label{fig:tutorialexample4}
	\end{minipage}}
	\subfigure[\href{https://stackoverflow.com/questions/14354885/android-fragments-backstack}{\textcolor{blue}{Fragments backStack}}]{
	\captionsetup{skip=0pt}
	\begin{minipage}{0.24\linewidth}
		\includegraphics[width=\linewidth]{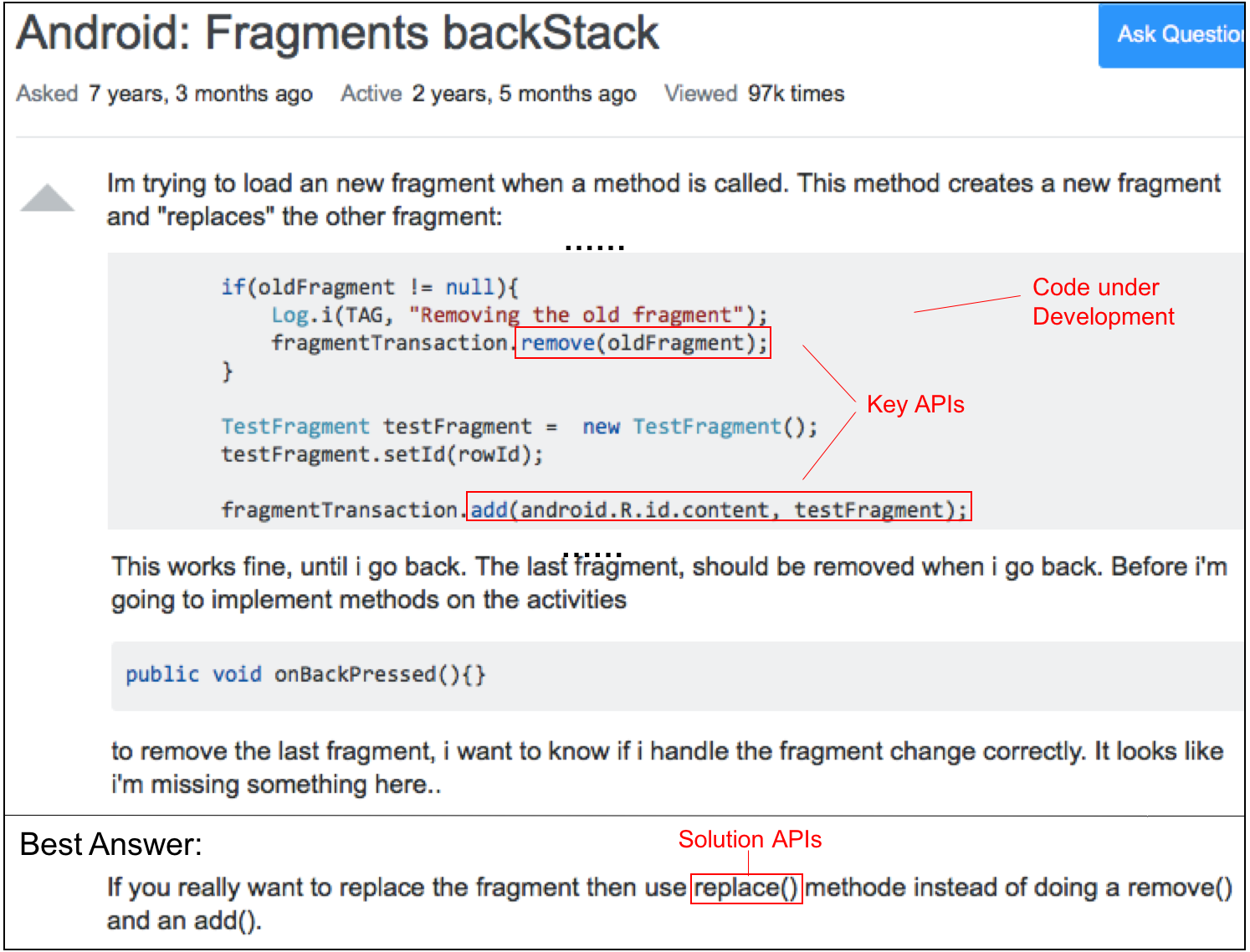}
		\vspace{-1mm}
		\label{fig:tutorialexample3}
	\end{minipage}}
	
	\vspace{-2mm}
	\caption{Examples of \textcolor{blue}{Stack Overflow Questions (blue)} and \textcolor{green}{Tutorials in Android Developer Guides (green)}}
	\vspace{-4mm}
	\label{fig:tutorialexamples}
\end{figure*}

To help API users, API developers often provide programming tutorials that document important programming tasks that their APIs support and demonstrate how to accomplish the tasks using code examples.
Figure~\ref{fig:tutorialexample2} and Figure~\ref{fig:tutorialexample4} shows the programming tutorials in \href{https://developer.android.com/guide}{\textcolor{green}{Android Developer Guides}} for two Android programming tasks.
The code examples in the tutorials illustrate the typical API usage scenarios for the tasks.
Such programming tutorials provide high-quality task-oriented knowledge to learn typical API usage scenarios.
However, the utilization of the provided tutorial knowledge in programming tasks is not satisfactory.

Sun et al.~\cite{taskkg} investigated 20 top-viewed Android how-to questions on Stack Overflow.
They found that for 16 out of these 20 questions, their accepted answers either excerpt or paraphrase the content in the tutorials in Android Developer Guides.
Unfortunately, for 8 of these 16 questions, the question askers tried to find solutions in the Android tutorials but failed. 
This difficulty was not just the problem of several question askers, but may potentially affect millions of developers (estimated by the large view count of these 16 questions).
We conduct a new formative study (see Section~\ref{sec:formativestudy}) to further understand the utilization of API usage examples in tutorials in real-world programming tasks and the barriers to effective use of API usage examples.
Our study involves 20 top-viewed Android-tagged questions (see Table~\ref{tab:formativestudy}), including 1 question from~\cite{taskkg} and 19 other questions, in which developers fail to use certain Android APIs to accomplish their tasks.
Two investigated questions (Q9/Q14) are shown in Figure~\ref{fig:tutorialexample1} and Figure~\ref{fig:tutorialexample3}.

Our study confirms the low utilization of API usage examples in Android Developer Guides by the question askers, even though Android Developer Guides contain effective code solutions to 19 investigated questions, as excerpted or referenced in the best answers of these questions.
We identify two major barriers to effective use of API usage examples.
First, \textit{task mismatch}.
The tutorials have only a set of programming tasks pre-defined by API developers, which is impossible to cover the API users' task needs in the wild.
For example, Android Developer Guides do not have an explicit task for ``create dialog without title'' that Q9 asks on Stack Overflow.
As shown in Figure~\ref{fig:tutorialexample1}, Q9 has been viewed 217k times (as of 30 April, 2020).
Second, \textit{code information overload}.
As illustrated in the code examples in Figure~\ref{fig:tutorialexample2} and Figure~\ref{fig:tutorialexample4}, the code examples often contain much irrelevant information, in addition to the code solutions to the specific issues that API users encounter, .

\vspace{0.5mm}
Our study also shows that document search (e.g., Google) can usually return relevant tutorial pages for the queries composed from the question titles, but, more often than not, it cannot identify the most relevant programming actions and code lines in (usually long) tutorial pages.
Sun et al.~\cite{taskkg} develop a programming task knowledge graph, which constitutes programming actions and actions relationships extracted from program tutorials, for example, the action ``performing fragment transactions'' in web page heading, the action ``preserve the previous state in the back stack'' in textual description, and the parent-child relation between the two actions in Figure~\ref{fig:tutorialexample4}.
They develop knowledge graph based task search by natural language queries.
Although promising in retrieving pertinent programming actions, our study shows that the current approach in~\cite{taskkg} has limitations in action coverage, natural language based search method and coarse-grained code example presentation, which make it ineffective in tackling the task mismatch and code information overload barriers.




\vspace{0.5mm}
We further examine the code snippets posted in the questions, the code solutions in the best answers, and the most relevant programming actions and code lines in the tutorials referenced in the best answers.
We find that the knowledge graph approach can overcome the task mismatch and code information overload barriers, if we support the task search by matching the code under development and the code examples in the tutorials, and extend the knowledge graph with comment-level actions and code snippets.
Code matching can find relevant programming actions by the key APIs involved in the developer's code and the tutorial code examples, no matter how different the tasks are described by the developers and the tutorials, such as the Q3 and Q4 in Table~\ref{tab:formativestudy} and the tutorial page ``Load Large Bitmaps Efficiently''.
Comment-level actions and code snippets have two benefits.
First, they enrich the knowledge graph with more fine-grained actions and code snippets (see the examples in Figure~\ref{fig:tutorialexample2} and Figure~\ref{fig:tutorialexample4}).
This increases the likelihood to cover the API users' tasks.
Second, they can be used to highlight steps involved in recommended code examples, and thus mitigate code information overload (see an example in Figure~\ref{fig:ui}).

Inspired by our formative study results, we enhance the knowledge graph construction method in~\cite{taskkg} from two perspectives.
First, we extract programming actions from comments in code examples in tutorials.
Second, we link the duplicate actions from code comments and textual description by a new duplication-action relationship, such as ``add the transaction to the back stack'' in code comment and ``preserve the previous state in the back stack'' in the text preceding the code example in Figure~\ref{fig:tutorialexample4}.
As a proof-of-concept, we construct a programming task knowledge graph from \href{https://developer.android.com/guide}{\textcolor{green}{Android Developer Guides}}.
This knowledge graph contains 23,918 programming actions and 4,050 code-example snippets.
To advance the utilization of API usage examples in tutorials, we integrate the enhanced programming task knowledge graph in the IntelliJ IDE.
We develop an IDE plugin (called CueMeIn) that saliently analyzes the code in a method under editing, and prompt relevant programming actions and code examples in tutorials to the developer.
Figure~\ref{fig:ui} shows a screenshot of our tool's recommendation for the code in the question in Figure~\ref{fig:tutorialexample1}.
The developer can explore the recommended actions and their parent/children/sibling actions in the Task Hierarchy panel, and inspect the tutorial excerpts and code examples of the programming actions (with key action and API information highlighted) in the Task Details panel.


By a statistical sampling method~\cite{Singh1996}, we confirm that the accuracy of various extracted action information is 79\% or above.
To evaluate the effectiveness of our API usage examples prompting, we investigate if our tool can recommend code examples for solving the 20 Stack Overflow questions in our formative study, given the code snippets posted in these questions. 
Our results show that for 15 out of the 20 questions, our tool recommends the code examples that contain effective code solutions mentioned in the best answers to the questions.
To validate the usefulness of our tool, we conduct a user study involving 12 Master students recruited in our school and 6 bug-fixing tasks derived from Stack Overflow questions. 
The user study results show that our tool helps developers find and use relevant code examples in tutorials faster and more effectively, compared with regular IDE support and document search.

In this work, we make following contributions: 

\begin{itemize}
	\item We conduct a formative study to understand the utilization of API usage examples in tutorials, the barriers to effective use and the limitation of existing search methods.
	
	\item We address the identified barriers by an enhanced programming task knowledge graph, the code-matching based task search, and the prompting of task-oriented API usage examples in the IDE based on the knowledge graph.
	
	\item We evaluate the quality of the constructed knowledge graph and the effectiveness and usefulness of our tool.
\end{itemize}

\section{Formative Study}
\label{sec:formativestudy}

\begin{table*}
	\small
	\centering
	\caption{Formative Study Results (RQ1 and RQ2) (Refer to Section~\ref{sec:questionselection} and Section~\ref{rq1rq2analysis} for the explanation of acronyms)}
	\label{tab:formativestudy}
	\begin{tabular}{lr}
		\toprule[1pt]
		\textbf{Stack Overflow Question} & \textbf{Tutorial Page Linked in Best Answer}\\
		\midrule
		\href{https://stackoverflow.com/questions/3500197}{1.\textcolor{blue}{How to \textbf{display Toast} in Android?}} [Attempt=No] &  \href{https://developer.android.com/guide/topics/ui/notifiers/toasts.html}{\textcolor{green}{Toasts overview}} [KS=Mod, 6RCL, 17TCL, 2RCB, 4TCB]\\
		
		\href{https://stackoverflow.com/questions/3875184}{2.\textcolor{blue}{Can't \textbf{create handler inside thread} that  has not called ...}} [Attempt=No]& \href{https://developer.android.com/training/multiple-threads/communicate-ui.html}{\textcolor{green}{Communicate with the UI}} [KS=Same, 17RCL, 96TCL, 4RCB, 6TCB]\\
		
		\href{https://stackoverflow.com/questions/477572}{3.\textcolor{blue}{Strange \textbf{out of memory} issue while loading  an image to a ...}} [Attempt=No]& \href{https://developer.android.com/topic/performance/graphics/load-bitmap}{\textcolor{green}{Loading Large Bitmaps}} [KS=Same, 18RCL, 32TCL, 3RCB, 4TCB]\\
		
		\href{https://stackoverflow.com/questions/32244851}{4.\textcolor{blue}{Android:java.lang.\textbf{OutOfMemoryError}:Failed to allocate a ...}} [Attempt=No] & \href{https://developer.android.com/topic/performance/graphics/load-bitmap}{\textcolor{green}{Loading Large Bitmaps}} [KS=Alter, 18RCL, 32TCL, 3RCB, 4TCB]\\

		\href{https://stackoverflow.com/questions/2183962}{5.\textcolor{blue}{How to \textbf{read value from string.xml} in android}} [Attempt=No]& \href{https://developer.android.com/guide/topics/resources/string-resource.html}{\textcolor{green}{String resources}} [KS=Same, 5RCL, 76TCL, 5RCB, 12TCB]\\

		\href{https://stackoverflow.com/questions/2614719}{6.\textcolor{blue}{How do I \textbf{get the SharedPreferences} from a PreferenceActi ...}} [Attempt=Yes] & \href{https://developer.android.com/training/data-storage#pref}{\textcolor{green}{Data and file storage overview}} [KS=BG, 0RCL, 0TCL, 0RCB, 0TCB]\\
		
		\href{https://stackoverflow.com/questions/5042197}{7.\textcolor{blue}{ \textbf{set height and width of Custom view programmatically}}} [Attempt=No] & \href{http://developer.android.com/guide/topics/ui/custom-components.html#custom}{\textcolor{green}{Fully Customized Components}} [KS=BG, 0RCL, 0TCL, 0RCB, 0TCB]\\
		
		\href{https://stackoverflow.com/questions/19451715}{8.\textcolor{blue}{Same \textbf{Navigation Drawer} in different Activities}} [Attempt=Yes]& \href{https://developer.android.com/guide/navigation/navigation-ui#add\_a\_navigation\_drawer}{\textcolor{green}{Add a navigation drawer}}   [KS=Alter, 10RCL, 90TCL, 2RCB, 12TCB]\\
		
		\href{https://stackoverflow.com/questions/2644134}{9.\textcolor{blue}{ How to create a \textbf{Dialog without a title}?}} [Attempt=No]& \href{https://developer.android.com/guide/topics/ui/dialogs.html#CustomDialog}{\textcolor{green}{Dialogs}}  [KS=Mod, 1RCL, 176TCL, 1RCB, 3TCB]\\

		\href{https://stackoverflow.com/questions/8330276}{10.\textcolor{blue}{\textbf{Write a file in external storage} in Android}} [Attempt=No]& \href{https://developer.android.com/training/data-storage#filesExternal}{\textcolor{green}{Data and file storage overview}} [KS=BG, 0RCL, 0TCL, 0RCB, 0TCB]\\
		
		\href{https://stackoverflow.com/questions/10285047}{11.\textcolor{blue}{\textbf{showDialog deprecated}. What's the alternative?}} [Attempt=Yes]& \href{https://developer.android.com/guide/components/fragments}{\textcolor{green}{Fragments}}[KS=BG. 0RCL, 152TCL, 0RCB, 14TCB]\\

		\href{https://stackoverflow.com/questions/25719620}{12.\textcolor{blue}{How to solve \textbf{java.lang.OutOfMemoryError}  trouble in ...}} [Attempt=No]& \href{https://developer.android.com/topic/performance/memory.html}{\textcolor{green}{Manage your app's memory}}  [KS=Mod, 24RCL, 31TCL, 2RCB, 2TCB]\\

		\href{https://stackoverflow.com/questions/16898675}{13.\textcolor{blue}{How it work -  \textbf{requestLocationUpdates()} + LocationRequest ...}} [Attempt=Yes]& \href{https://developer.android.com/training/location/index.html}{\textcolor{green}{Build location-aware apps}}  [KS=BG, 0RCL, 0TCL, 0RCB, 0TCB] \\
		
		\href{https://stackoverflow.com/questions/14354885}{14.\textcolor{blue}{Android: \textbf{Fragments} backStack}}  [Attempt=No]& \href{https://developer.android.com/guide/components/fragments.html}{\textcolor{green}{Fragments}} [KS=Same, 5RCL, 152TCL, 1RCB, 14TCB]\\
		
		\href{https://stackoverflow.com/questions/6369103}{15.\textcolor{blue}{how to \textbf{set imageview src}?}} [Attempt=No] & \href{https://developer.android.com/guide/topics/resources/providing-resources#ResourcesFromCode}{\textcolor{green}{Accessing resources in code}}  [KS=Same, 2RCL, 2TCL, 1RCB, 1TCB]\\

		\href{https://stackoverflow.com/questions/29273387}{16.\textcolor{blue}{\textbf{CertPathValidatorException} : Trust anchor for certificate ...}} [Attempt=No]& \href{https://developer.android.com/training/articles/security-ssl.html#UnknownCa}{\textcolor{green}{Unknown certificate}} [KS=Same, 24RCL, 24TCL, 1RCB, 1TCB]\\
		
		\href{https://stackoverflow.com/questions/2542938}{17.\textcolor{blue}{SharedPreferences.\textbf{onSharedPreferenceChangeListener} ...}} [Attempt=No] & \href{https://developer.android.com/guide/topics/ui/settings.html#Listening}{\textcolor{green}{Settings}} [KS=BG, 0RCL, 16TCL, 0RCB, 12TCB]\\
		
		\href{https://stackoverflow.com/questions/7618703}{18.\textcolor{blue}{Activity lifecycle - \textbf{onCreate called on every re-orientation}}} [Attempt=No]& \href{https://developer.android.com/guide/topics/resources/runtime-changes.html}{\textcolor{green}{Handle configuration}} [KS=Alter, 7RCL, 9TCL, 1RCB, 1TCB]\\

		\href{https://stackoverflow.com/questions/13305861}{19.\textcolor{blue}{Fool-proof way to \textbf{handle Fragment on orientation change}}} [Attempt=No] & \href{https://developer.android.com/guide/components/fragments.html}{\textcolor{green}{Fragments}}  [KS=Mod, 14RCL, 152TCL, 1RCB, 14TCB]\\

		\href{https://stackoverflow.com/questions/4169714}{20.\textcolor{blue}{How to \textbf{call Activity from a menu item} in Android?}} [Attempt=No]& \href{https://developer.android.com/guide/topics/ui/menus.html}{\textcolor{green}{Menus}} [KS=Mod, 21RCL, 62TCL, 2RCB, 11TCB]\\
		\bottomrule[1pt]
	\end{tabular}
	\vspace{-4mm}
\end{table*}

We conduct a formative study of Stack Overflow questions to investigate the following three research questions:

\begin{itemize}[leftmargin=*]
	\item \textbf{RQ1:} Can developers effectively find and use API usage examples in tutorials?
	
	\item \textbf{RQ2:} What are the barriers to the effective use of API usage examples in tutorials?
	
	\item \textbf{RQ3:} Can document search or activity-centric search~\cite{taskkg} overcome these barriers?
	
\end{itemize}

\subsection{Study Setup}
We select 20 Stack Overflow questions for detailed analysis.
One author performs the analysis as described below, and the other author validates the analysis results.
The disagreements are discussed by all co-authors to reach the consensus.

\subsubsection{Question Selection}\label{sec:questionselection}
The Question column in Table~\ref{tab:formativestudy} summarizes the 20 selected questions.
Due to the space limitation, we show only some keywords in question titles.
Readers can click the links to visit the questions on Stack Overflow.
In this work, we limit this study to android-tagged questions.
As our study focuses on API usage issues and solutions, we consider only the questions whose question bodies contain non-XML code fragments and optionally stack traces. 
Furthermore, to investigate the use of tutorial knowledge in solving API usage issues, we examine the questions whose best (i.e., accepted or most up-voted) answers contain URL link(s) to some web page(s) in Android Developer Guides.
We examine the questions by the descending order of their view counts.
We read the question and answer contents and select the top-20 ranked questions that discuss Android API usage issues and solutions.

\subsubsection{The Analysis to Answer RQ1 and RQ2}\label{rq1rq2analysis}
The analysis results are summarized in the columns Question and Tutorial Page Linked in Best Answer in Table~\ref{tab:formativestudy}.
For each question, we check if the question asker describes some attempt to solve the issue with the knowledge in Android Developer Guides (Attempt=Yes or No), for example, by referencing some tutorial page content in the question, or mentioning some search of tutorial website.

We visit each tutorial page linked in the best answer of the question.
For each linked page, we count the number of non-XML code blocks (Total Code Block or TCB) (tagged by $<$devsite-code$>$ in Android Developer Guides) in the page and the total number of non-comment code lines (Total Code Line or TCL) in code blocks. 
We examine the Knowledge Support (KS) provided by the tutorial page, specifically, if the tutorial page contains some code examples that can solve the issue in the question  (e.g., the examples in Figure~\ref{fig:tutorialexamples}), or contains only some background knowledge relevant to the issue.
In the former case, we further examine if the solution in the best answer is the same as (KS=Same), modified from (KS=Mod) or an alternative to (KS=Alter) the code solution in the tutorial page.
We count the number of non-XML code blocks in the page that contain the code solution (Related Code Block or RCB) and the number of non-comment code lines related to the code solution (Related Code Line or RCL).
In the latter case, we mark KS=BG.

\vspace{2mm}
\subsubsection{The Analysis to Answer RQ3}
\label{rq3analysis}
The analysis results are presented in the columns Google Search and Activity-Centric Search in Table~\ref{tab:formativestudy-rq3}.
For each question, we select some keywords in the question title, such as those in bold font in Table~\ref{tab:formativestudy}.
We formulate queries using the selected keywords and their synonyms (e.g., ``dialog without title'' and ``remove dialog title'' for Q9) for Google search and activity-centric search.
For each question, we trial up to three queries and record the best search result.

We use Google to search the pages in the Android Developer Guides.
If a tutorial page in the top 10 search results is linked in the best answer, we consider Google can find relevant page (RelPage) to the question.
We further examine if the search result snippet contains solution-specific content excerpt from the relevant tutorial page.
If so, we consider Google can find solution-specific hints (SSHint).
We also search the original programming task knowledge graph by activity-centric search developed in~\cite{taskkg}.
Activity centric search returns answer snippets consisting of relevant programming actions and code examples to the query.
We examine the top-10 answer snippets: if an answer snippet contains solution-specific programming action and code example (SSAction), or just contains some relevant programming action (RelAction).
NoRel indicates no relevant actions in the top-10 answer snippets.

\subsection{RQ1: Utilization of API Usage Examples}
\label{sec:formativestudy-rq1}
Only 4 question askers (Q6/Q8/Q11/Q13) described some attempt to find solutions to their programming issues in Android Developer Guides.
As their attempts failed to find solutions, they asked the questions on Stack Overflow.
However, there is only one question (Q11) that really does not have the solution in Android Developer Guides.
Q11 asks for the alternative to an deprecated API.
This type of API knowledge is usually described in API reference, but not in programming tutorials.
For the rest 19 questions, their solutions can all be found in Android Developer Guides.

In fact, for 6 of these 19 questions (Q2/Q3/Q5/Q14/Q15/Q16), the best answers simply excerpt the code solutions in the referenced tutorial pages (KS=Same).
For 5 questions (Q1/Q9/Q12/Q19/Q20), the best answers modify the code solutions in the referenced tutorial pages (KS=Mod) to fit the question contexts.
For 3 questions (Q4/Q8/Q18), the best answers reference the tutorial pages that provide alternative solutions (KG=Alter) to the solutions in the answers.
For 5 questions (Q6/Q7/Q10/Q13/Q17), although the tutorial pages referenced in the best answers provide only background knowledge (KS=BG), our CueMeIn tool finds code examples in other tutorial pages (see Table~\ref{tab:formativestudy-rq3}) that can solve Q6/Q10/Q13/Q17. 

The 20 questions have minimum view count 76k, maximum 1 million and median 148k.
That is, a large number of developers may encounter the similar issues.
Although Stack Overflow provides a proxy to code examples in tutorials to fix these issues, the direct utilization of the code examples in tutorials is low.
In fact, some answerers point out this low utilization issue.
For example, one user comments on Q3 that ``This happens when you don't read the Android developer guides''.

\begin{table*}
	\small
	\centering
	\caption{Formative Study Results (RQ3) (Refer to Section~\ref{rq3analysis} for the explanation of acronyms) and Recommendations by Our CueMeIn Tool (Refer to Section~\ref{sec:effect} for the explanation of acronyms)}
	\label{tab:formativestudy-rq3}
	\small
	\begin{tabular}{lrrr}
		\toprule[1pt]
		\textbf{Ques-}  & \textbf{Google} & \textbf{Activity}&\textbf{CueMeIn}\\
		\textbf{tions}& \textbf{Search} & \textbf{Search~\cite{taskkg}}&\textbf{Recommendation}\\
		\midrule
		\href{https://stackoverflow.com/questions/3500197}{\textcolor{blue}{Q1}}  & SS-Hint & SSAction&\href{https://developer.android.com/training/sign-in/biometric-auth#display-login-prompt}{\textcolor{green}{Display the login prompt}} [Rank-1, Input=AllCode, TaskRel=No, CommentAns=No]\\
		
		\href{https://stackoverflow.com/questions/3875184}{\textcolor{blue}{Q2}} &SSHint & SSAction & \href{https://developer.android.com/guide/components/aidl#Calling}{\textcolor{green}{Calling an IPC method}} [Rank-1, Input=KeyAPI, TaskRel=No, CommentAns=No]\\
		
		\href{https://stackoverflow.com/questions/477572}{\textcolor{blue}{Q3}} &RelPage& NoRel & \href{https://developer.android.com/training/camera/photobasics#TaskScalePhoto}{\textcolor{green}{Decode a scaled Bitmaps}} [Rank-1, Input=KeyAPI, TaskRel=Related, CommentAns=Rank-2]\\

		\href{https://stackoverflow.com/questions/32244851}{\textcolor{blue}{Q4}}  & RelPage & SSAction & \href{https://developer.android.com/training/camera/photobasics#TaskScalePhoto}{\textcolor{green}{Decode a scaled Bitmaps}} [Rank-2, Input=AllCode, TaskRel=Related, CommentAns=HL]\\

		\href{https://stackoverflow.com/questions/2183962}{\textcolor{blue}{Q5}} & RelPage & NoRel & \href{https://developer.android.com/guide/topics/connectivity/bluetooth-le?hl=en#read}{\textcolor{green}{Read BLE attributes}} [Rank-2, Input=AllCode, TaskRel=No, CommentAns=No]\\

		\href{https://stackoverflow.com/questions/2614719}{\textcolor{blue}{Q6}}  & SSHint & SSAction & \href{https://developer.android.com/training/basics/network-ops/xml#consume}{\textcolor{green}{Consume XML data}} [Rank-1, Input=KeyAPI, TaskRel=No, CommentAns=Rank-3]\\

		\href{https://stackoverflow.com/questions/5042197}{\textcolor{blue}{Q7}} & NoRel & NoRel & -\\

		\href{https://stackoverflow.com/questions/19451715}{\textcolor{blue}{Q8}}&RelPage& NoRel & -\\

		\href{https://stackoverflow.com/questions/2644134}{\textcolor{blue}{Q9}}& SSHint & NoRel & \href{https://developer.android.com/guide/topics/ui/dialogs.html#FullscreenDialog}{\textcolor{green}{Showing a Dialog Fullscreen}} [Rank-2, Input=AllCode, TaskRel=Subtask, CommentAns=HL]\\

		\href{https://stackoverflow.com/questions/8330276}{\textcolor{blue}{Q10}}& RelPage & NoRel & \href{https://developer.android.com/guide/topics/media/camera#capture-picture}{\textcolor{green}{Capturing pictures}} [Rank-2, Input=AllCode, TaskRel=No, CommentAns=No]\\

		\href{https://stackoverflow.com/questions/10285047}{\textcolor{blue}{Q11}} & RelPage& NoRel & -\\

		\href{https://stackoverflow.com/questions/25719620}{\textcolor{blue}{Q12}} & RelPage& NoRel & -\\

		\href{https://stackoverflow.com/questions/16898675}{\textcolor{blue}{Q13}} & SSHint & RelAction & \href{https://developer.android.com/training/location/request-updates#updates}{\textcolor{green}{Make a location request}} [Rank-1, Input=KeyAPI, TaskRel=Subtask, CommentAns=No]\\

		\href{https://stackoverflow.com/questions/14354885}{\textcolor{blue}{Q14}} & SSHint & SSAction & \href{https://developer.android.com/guide/navigation/navigation-migrate#single_activity_managing_multiple_fragments}{\textcolor{green}{managing multiple fragments}} [Rank-1, Input=AllCode, TaskRel=No, CommentAns=HL]\\

		\href{https://stackoverflow.com/questions/6369103}{\textcolor{blue}{Q15}} & SSHint & SSAction & \href{https://developer.android.com/guide/topics/graphics/drawables#drawables-from-images}{\textcolor{green}{Create drawables app}} [Rank-3, Input=AllCode, TaskRel=Related, CommentAns=No]\\

		\href{https://stackoverflow.com/questions/29273387}{\textcolor{blue}{Q16}} &  RelPage & NoRel & \href{https://developer.android.com/training/articles/security-ssl.html#UnknownCa}{\textcolor{green}{Unknown certificate}} [Rank-1, Input=AllCode, TaskRel=Same, CommentAns=Rank-2]\\

		\href{https://stackoverflow.com/questions/2542938}{\textcolor{blue}{Q17}} &SSHint & NoRel & \href{https://developer.android.com/training/basics/network-ops/managing#implement-preference-activity}{\textcolor{green}{Implement a preference }} [Rank-1, Input=AllCode, TaskRel=No, CommentAns=Rank-2]\\

		\href{https://stackoverflow.com/questions/7618703}{\textcolor{blue}{Q18}} & RelPage& NoRel & -\\

		\href{https://stackoverflow.com/questions/13305861}{\textcolor{blue}{Q19}} & RelPage& NoRel & \href{https://developer.android.com/guide/components/fragments.html#Example}{\textcolor{green}{Example}} [Rank-2, Input=AllCode, TaskRel=Subtask, CommentAns=HL]\\

		\href{https://stackoverflow.com/questions/4169714}{\textcolor{blue}{Q20}} & RelPage & NoRel & \href{https://developer.android.com/guide/topics/ui/menus#checkable}{\textcolor{green}{Using checkable menu}} [Rank-1, Input=AllCode, TaskRel=Subtask, CommentAns=No]\\
		\bottomrule[1pt]
	\end{tabular}
	\vspace{-5mm}
\end{table*}

\subsection{RQ2: Barrier to Effective Use}
We identify two major barriers to effective use of API usage examples: task mismatch and code information overload.

\subsubsection{Task Mismatch}
For the 19 questions that have code solutions in Android Developer Guides, there are only 6 questions whose question titles are somewhat similar to the descriptions of relevant programming actions in the tutorials, such as the title ``How to display Toast in Android'' of Q1 and the action ``Display the Toast'' in the tutorial page \href{https://developer.android.com/guide/topics/ui/notifiers/toasts.html\#Basics}{Toasts Overview}.

API developers cannot define tutorial tasks for all possible API usage scenarios.
For example, Android Developer Guides do not have a task for ``create dialog without title'' as Q9 needs.
As shown in Figure~\ref{fig:tutorialexample2}, the API usage for ``create dialog without title'' is part of the code example for the task ``show a dialog fullscreen ...'' which has no similarity to the Q9's title.
As another example, Q3 and Q4 have different task contexts: Q3 wants to load image into ListView, while Q4 wants to compress the image.
Despite of the different task contexts, the core issue of the two questions are the same, i.e., how to load large image efficiently.
Although the relevant code solution is discussed in the tutorial ``Load Large Bitmaps Efficiently'', it is not presented in the tasks similar to Q3 and Q4.

\subsubsection{Code Information Overload} 
Our analysis shows that there are usually multiple code blocks (TCB) in the referenced tutorial pages and these code blocks may contain tens or hundreds of non-comment code lines (TCL).
However, it is often the case that only a small number of code lines in some code blocks is relevant to solving a particular programming issue.
For the tutorial pages that have large numbers of code blocks and/or lines of code, such as Q2/Q5/Q8/Q9/Q14/Q19/Q20, it is not easy to spot such most relevant code.
For example, the referenced tutorial page Dialogs for Q9 has 3 code blocks with 176 lines of code.
But only 1 line of code in 1 code block is the code solution to the Q9 (see Figure~\ref{fig:tutorialexample2}).
In fact, the value of Stack Overflow answers is to distill such most relevant code solutions for the questions.
For example, for 11 of the 20 questions we investigate, their best answers excerpt the important information from the tutorial, summarize key programming actions, and show the key code lines with concise description. 

\subsection{RQ3: How to Overcome the Barriers?}

As shown in Table~\ref{tab:formativestudy-rq3}, Google search finds relevant tutorial pages (RelPage) for 19 questions.
For 8 of these 19 questions, the search result snippet of the relevant page contains solution-specific content excerpt (SSHint).
For example, Google returns the \href{https://developer.android.com/guide/topics/ui/dialogs.html}{\textcolor{green}{Dialog}} page for the query ``dialog without title'' for Q9.
The search result snippet includes the code comment ``... you can remove the dialog title ...'' in the code example in the Dialog page.
Such solution-specific content could mitigate code information overload.
However, developers still need to manually search the solution-specific content in the tutorial page in order to find the code solution in the page.
For the other 11 questions, developers have to read lengthy tutorial pages to find relevant code solutions in them.

Activity-centric search over knowledge graph~\cite{taskkg} finds programming actions that contain code solutions for 6 questions (SSAc-tion), and relevant programming action for one question (RelAction).
The failure of activity-centric search for the rest 13 questions is because the underly knowledge graph does not contain relevant programming actions.
We find that the action coverage can be improved by adding comment-level actions in code examples to the knowledge graph.
Some actions in code comments, such as ``add the transaction to the back stack'' in Figure~\ref{fig:tutorialexample4} duplicate actions explained in text, but many other actions, such as ``remove dialog title'' in Figure~\ref{fig:tutorialexample2} and ``create new fragment and transaction'' in Figure~\ref{fig:tutorialexample4} represents fine-grained API usage scenarios that have not been explicitly defined as programming tasks in the tutorials.
Another improvement is to find code solutions by matching the code under development and the code examples in programming tutorials.
This is because code matching relies on the key APIs involved in the developer's code and the tutorial code examples, rather than natural language descriptions of programming tasks.
 
Another important limitation is the search results presentation.
In the current knowledge graph, a whole code example is linked to an action or a set of sibling actions immediately preceding the code example.
When there is no such preceding action(s), it links the code example to the parent action.
Although this heuristic is effective in linking code examples to relevant actions, it does not consider the smaller code fragments in a code example, which often correspond to more fine-grained API usage scenarios.
As shown in Figure~\ref{fig:tutorialexamples}, it is often the case that only some parts of the code examples of relevant programming actions are relevant to the developer's need.
However, it may not be easy to spot the most relevant parts of the code examples, especially when the code examples are long.

\noindent
\fbox{
	\begin{minipage}{8.2cm} \emph{The utilization of API usage examples in the tutorials is low due to two major barriers: task mismatch and code information overload. Although document search is effective in finding relevant documents, it cannot effectively find programming actions and code examples. Although the knowledge graph supports direct search of programming actions, the existing knowledge graph approach~\cite{taskkg} has limitations in action coverage, natural language based search method, and coarse-grained code example recommendation.}
\end{minipage}}

\section{Approach}
To address the limitations of existing document search and activity-centric search~\cite{taskkg}, we enhance the knowledge graph with more actions and finer-grained code fragments (Section~\ref{sec:enhancingtaskkg}), support task search by matching the code under development and the code examples in tutorials (Section~\ref{sec:codematching}), and prompt the developer with informative programming actions and code examples relevant to the code under development (Section~\ref{sec:prompting}).

\subsection{Enhancing Knowledge Graph}
\label{sec:enhancingtaskkg}

\begin{figure}
	\centering
	\includegraphics[scale=0.52]{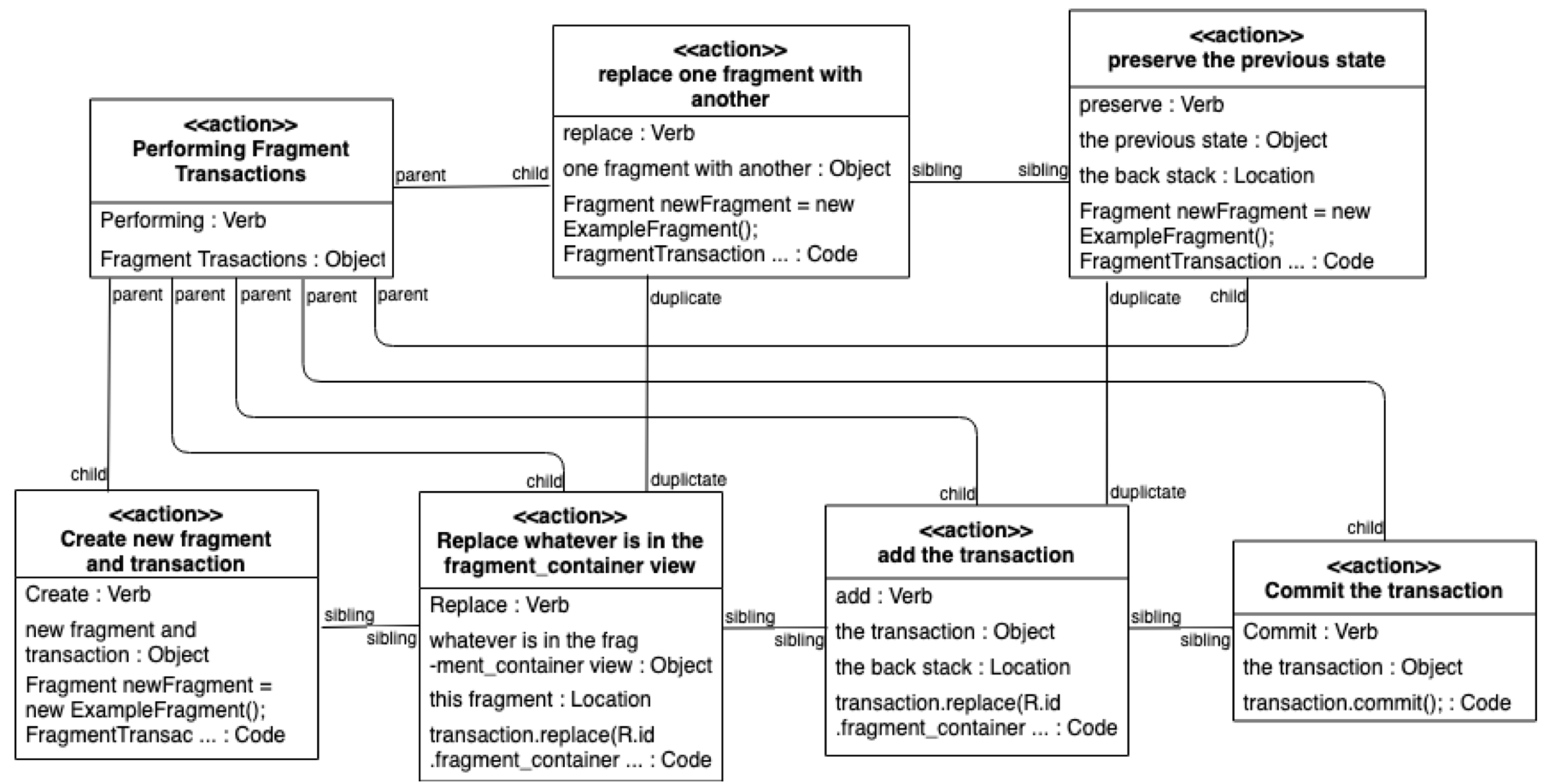}
	\caption{Knowledge Graph for the Tutorial Page in Figure~\ref{fig:tutorialexample4}}
	\label{fig:kgexample}
	\vspace{-6mm}
\end{figure}

Our approach extends the programming task knowledge graph and the knowledge graph construction method in~\cite{taskkg}.

\subsubsection{Knowledge Graph Schema and Construction Method}
As illustrate in Figure~\ref{fig:kgexample}, in the programming task knowledge graph proposed in~\cite{taskkg}, entities are \textit{programming actions}.
An action can be decomposed into sub-actions, which forms an action hierarchy.
In addition to the \textit{hierarchical (i.e., parent-child )} relationships between an action and its sub-actions, there are two more types of action relationships: 
\textit{descriptive sibling} (a sibling action mentioned immediately before another sibling action), and \textit{preceding-following} (sibling actions with clear step indicators).
An action has a verb phrase including an \textit{action verb} and an \textit{object}, and may have the following attributes: 1) \textit{API} involved in the action; 2) \textit{location} where the action occurs; 3) \textit{condition} to satisfy for performing the action; 4) \textit{goal} that the action achieves; 5) \textit{code} demonstrating the action.
 
The construction method include: \textit{action extraction}, \textit{attribution extraction} and \textit{relation extraction}.
As annotated in Figure~\ref{fig:tutorialexample4}, actions are extracted from document headings and textual descriptions.
To overcome the rigidity of pre-defined verb and noun lists~\cite{TaskNav15}, a sentence classifier is trained to classify a sentence as activity or non-activity sentence.
An activity sentence is parsed to obtain Part-of-Speech (POS) tags using natural language processing tools (e.g., CoreNLP~\cite{manning-corenlp2014} or Spacy~\cite{honnibal-johnson2015}).
Action verb phrase are then extracted from the sentence based on the POS tag patterns developed in~\cite{zhaoxuejiaosaner16}, APIs and code snippets are extracted by the tags used to annotate APIs and code snippets in the tutorial documents, and other action attributes are extracted keyword patterns.
Hierarchical relationships are extracted from document structures, precede-follow relationships are determined by the explicit step indicators (e.g., $<$ol$>$ tag), and descriptive sibling relationships are determined by the actions' mention order in the sentence.

\vspace{3mm}
\subsubsection{Our Extensions}
We adopt the original knowledge graph schema and add one more type of action relationship - \textit{duplicate action}.
We adopt the original knowledge graph construction method and make the following three extensions to improve action coverage and increase fine-grained code snippets in the knowledge graph.

First, we consider comments in code examples as an additional source of programming actions.
Code comments are more action intensive than tutorial text, because code comments usually explain what some code is intended to do.
In contrast, tutorial text often has to explain much background knowledge beyond the actions to do.
One issue with the actions in code comments is that they may repeat the actions already explained in tutorial text, but often in different ways, such as the two pairs of duplication actions in Figure~\ref{fig:tutorialexample4}.
We train a BERT-based duplicate sentence classifier to detect such duplicate action sentences.
The classifier takes as input one action sentence from the comment and one action sentence from tutorial text under the same parent action, and predict whether the two sentences are duplicate.
If so, we add a duplicate relation between the two actions, and link the code snippet of the comment action to the duplicate action from tutorial text.

Second, we expand the form of activity sentences and the training data for activity-sentence classifier.
Sun et al.~\cite{taskkg} considers only imperative sentences as activity sentences.
However, we find that many non-imperative sentences also describe programming actions, often in the form of ``you can/need/must ...'', for example ``you can remove the dialog title, but you must call the superclass to get the Dialog''.
However, not all sentences in this form are activity sentences, for example ``you can learn more about the other app components''.
We expand the training data for the activity-sentence classifier, originally having 20,560 activity and 20,786 non-activity sentences, with 2,250 activity and 3,594 non-activity sentences in the new forms from tutorial text.
We also expand the training data with 230 activity and 230 non-activity sentences from code comments, including both imperative sentences and new forms.

Third, we replace the BiLSTM (Bi-directional Long Short Term Memory) based activity sentence classifier with the BERT-based classifier.
BERT (Bidirectional Encoder Representations from Transformers~\cite{bert2018}) is the state-of-the-art sentence embedding model.
Furthermore, the self-attention mechanism of BERT could learn better embeddings in the presence of co-references in the sentence, such as ``one fragment'' and ``previous state'' in ``replace one fragment with another, and preserve the previous state to the back stack'', or ``layout'' and ``it'' in ``create a layout and add it to an AlertDialog''.

\subsection{Code Matching based Task Search}
\label{sec:codematching}

Different from activity search by natural language queries in~\cite{taskkg}, our approach recommends relevant programming actions by matching the code under development and the code examples associated with programming actions.
The underlying rationale is that programming tasks may be described in different ways, but the key APIs involved in the tasks and the programming issues developers may encounter could be the same or similar.
Therefore, code matching based task search helps to overcome the task mismatch barrier faced by activity search by natural language queries.

In this work, we consider the code in a method currently being edited as the code under development, denoted as $C_D$.
We denote a code snippet associated with a programming action in the knowledge graph as $C_A$.
As both $C_D$ and $C_A$ can be partial or incompilable code, we use the API recognition method proposed in~\cite{liveAPIdocicse2014} to identify the APIs used in the code.
First, we crawl the APIs from the official API reference website, for example, \href{https://developer.android.com/reference/packages}{Android API reference} for Android APIs, and build an API dictionary.
Given a code snippet, we first selects a set of candidate APIs in the API dictionary whose name matches some name in the code snippet.
We then use the context-based disambiguation mechanism proposed in~\cite{liveAPIdocicse2014} to determine the most likely APIs used in the code snippet.
This context-based disambiguation mechanism essentially determines unique APIs by filtering irrelevant candidate APIs based on code elements co-occurring in the same file of the code snippet. 

Let $API_{C_D}$ and $API_{C_A}$ be the set of APIs used in the code under development $C_D$ and the code snippet in knowledge graph $C_A$ respectively.
The APIs can be classes, methods, fields and/or constant values.
The similarity of $C_D$ and $C_A$ is measured by $sim(C_D, C_A)=(\lambda_1*Match(API_{C_D}, API_{C_A})+\lambda_2*Unmatch(API_{C_D}, API_{C_A}))/|API_{C_A}|$.
$Match(API_{C_D}, API_{C_A})$ counts the number of APIs $a \in API_{C_A}$ that are used in $API_{C_D}$.
$Unmatch(API_{C_D}, API_{C_A})$ counts the number of APIs $a \in API_{C_A}$ that are not used in $API_{C_D}$.

There are three variations points in this similarity computation.
First, we can let $API_{C_D}$ (or $API_{C_A}$) contain specific APIs used in the code, or abstract API methods/fields/constants to their declaring classes.
Using declaring classes of specific APIs would allow for more generality for code matching, but may result in less specific matching results.
Second, we can let $API_{C_A}$ be a set or a multiset (i.e., bag).
That is, we can consider just whether an API is used or not in the code or how many times an API is used.
API usage times could be an indicator of the importance of an API in the code.
Third, we can consider only matched APIs ($\lambda_2 = 0$ in this case) or both matched and unmatched APIs.
When searching for API usage examples, considering unmatched APIs in $API_{C_A}$ would allow for serendipitous discovery of API usage scenarios that developer may not be aware of.
However, we would like to guarantee certain degree of matching between the code under development and the code examples, while allowing for this serendipitous discovery.
Therefore, we give matched APIs more weight than unmatched APIs.
In this work, we set $\lambda_1=2$ and $\lambda_2=1$.

\begin{figure}
	\centering
	\includegraphics[width=\linewidth]{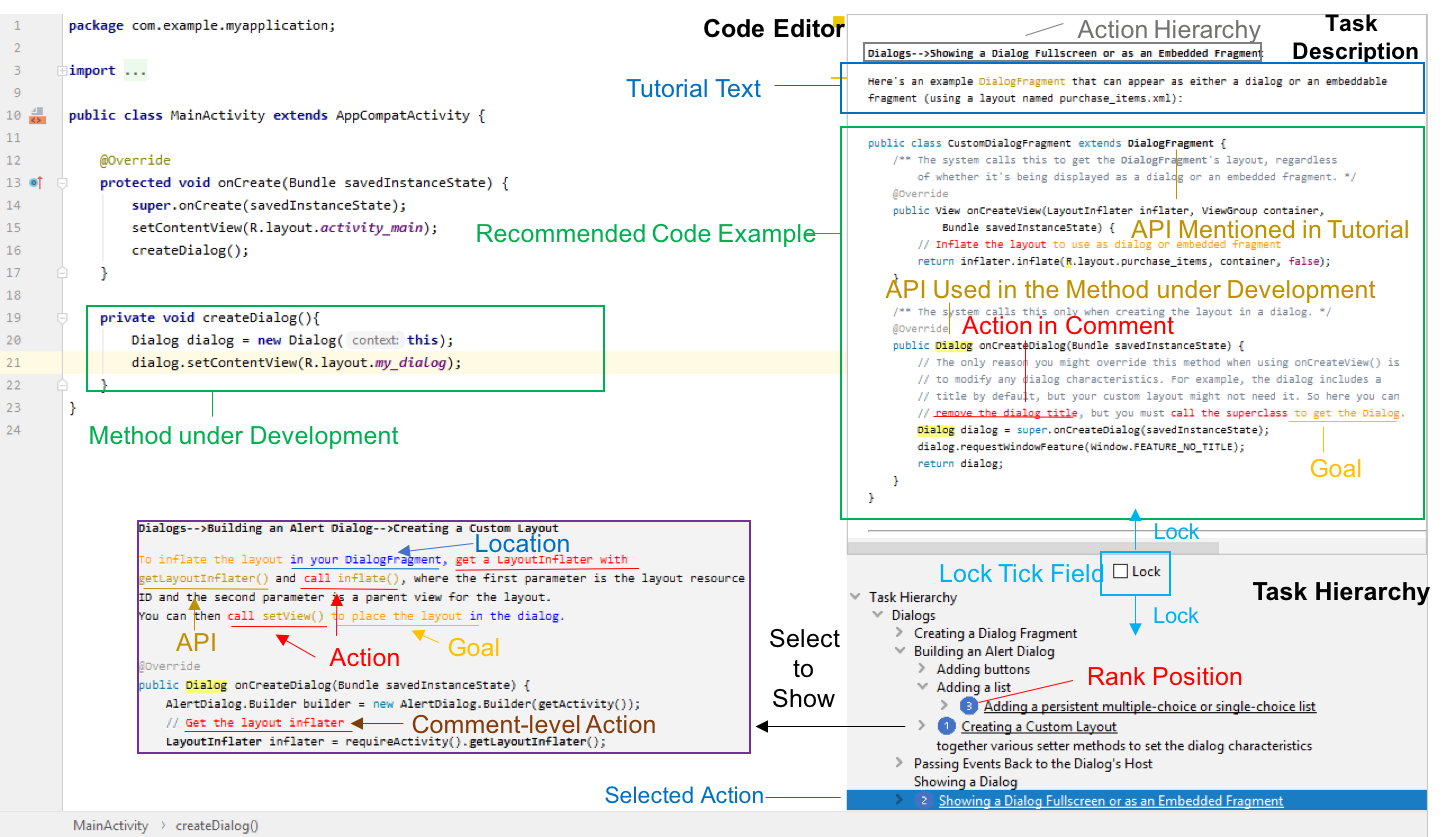}
	\vspace{-2mm}
	\caption{The User Interface of Our CueMeIn Tool}
	\label{fig:ui}
	\vspace{-6mm}
\end{figure}

\vspace{2mm}
\subsection{In-IDE API Usage Example Prompting}
\label{sec:prompting}

Different from the query-response paradigm in~\cite{taskkg}, we adopt an observe-push paradigm to prompt developers with API usage examples while they are programming the code.
This observe-push paradigm can be integrated into the development environment.
As a proof of concept, we implement it in a plugin (called CueMeIn) of the IntelliJ IDE.
Figure~\ref{fig:ui} shows a screenshot of our plugin's user interface.
CueMeIn saliently observes and analyzes the code in a method that the developers is editing.
When the developer pauses the editing longer than an interval (5 second in current tool), CueMeIn matches the code in the method under development with the code snippets in the knowledge graph as described in Section~\ref{sec:codematching}.
The developer can also select some code lines or elements, which also trigger the code matching.
CueMeIn recommends the top-N (N=3 in the current tool) code snippets with the highest scores and the corresponding programming actions.

The recommended programming actions are shown in the Task Hierarchy panel.
CueMeIn does not simply show the recommended programming actions in a list.
Instead, it shows the ancestor actions of a recommended action, up to the top-level task.
In addition, it shows other children actions of this top-level task.
To avoid information overload, CueMeIn expands only the hierarchical path from the top-level task to the recommended programming action, but it keeps all other actions collapsed.
The recommended actions are underlined and annotated with the rank positions.
The developer can expand or collapse any action nodes to explore the task hierarchy.

When a programming action is selected, its corresponding tutorial excerpt will be displayed in the Task Description panel.
The action hierarchy is shown at the top of the panel.
If the tutorial excerpt contains the recommended code snippets, this recommended code snippet is shown in gray background.
This helps to distinguish the recommended code snippet from other code blocks in the tutorial description. 
To help the developer read and spot important information, such as programming actions, APIs, goals, locations in the task description, CueMeIn highlights action phrases and attributes in the description based on the knowledge graph. 
It shows in bold font the APIs in the tutorial code snippets that are mentioned in the tutorial text (e.g., DialogFragment).
This helps the developer spot relevant explanation of API usage.
The plugin highlights in yellow background the APIs in the tutorial code snippets that are used in the code under development.
This helps the developer understand the correspondence between the code under development and the code snippets in the tutorial.

In Figure~\ref{fig:ui}, the code in createDialog() is under development, corresponding to the code snippet in the question ``how to create a dialog without a title''  in Figure~\ref{fig:tutorialexample1}.
CueMeIn recommends three relevant code examples associated with three programming actions.
The code solution that the question asker seeks for is in the code example of the rank-2 action ``showing a dialog full screen or as an embedded fragment''.
Although this action itself is irrelevant to the issue the question asker tries to solve, the code example of this action contains API usage example which can solve the issue.
The code solution is associated with the action ``remove the dialog title'' in the code comment.
The action highlight can help the developer spot this code solution buried in other irrelevant code lines.

\section{Evaluation}
We conduct a set of experiments to investigate the following three research questions:

\begin{itemize}[leftmargin=*]
	\item \textbf{RQ4}: What is the accuracy of action/attribute/relation extraction for constructing knowledge graph?
	\item \textbf{RQ5}: How effective can code matching based task search recommend code solutions to the programming questions?
	\item \textbf{RQ6}: How useful is our CueMeIn tool for helping developers accomplish programming tasks?
\end{itemize}

\subsection{RQ4: Accuracy of Information Extraction}

We confirm the accuracy of our information extraction methods.

\subsubsection{Accuracy of Activity Sentence Classification}

\begin{table}
	\small
	\renewcommand\arraystretch{1.5}
	\centering
	\caption{Performance of Activity Sentence Classification}
	\label{tab:sentenceclassification}
	\begin{tabular}{cccc}
		\toprule
		\textbf{Method}&\textbf{Precision}&\textbf{Recall}&\textbf{F1 Score}\\
		\hline
		BiLSTM (Original Data)&0.93&0.97&0.95\\
		\hline
		BERT (Original Data)&\textbf{0.97}&\textbf{0.99}&\textbf{0.98}\\
		\hline
		BiLSTM (Enriched Data)&0.94&0.97&0.95\\
		\hline
		BERT (Enriched Data)&\textbf{0.97}&\textbf{0.99}&\textbf{0.98}\\
		\bottomrule
	\end{tabular}
	\vspace{-6mm}
\end{table}

As discussed in Section~\ref{sec:enhancingtaskkg}, we enrich the original dataset for activity sentence classifier with new forms of activity sentences and sentences from code comments.
We also replace the original BiLSTM-based classifier with the state-of-the-art BERT-based classifier.
We perform 10-fold cross-validation to confirm the accuracy of activity sentence classification.
As shown in Table \ref{tab:sentenceclassification}, a classifier's performance on the original dataset and the enriched dataset has no difference.
On both datasets, BERT-based classifier outperforms BiLSTM-based classifier.
The differences of prediction results  between BERT and BiLSTM is statistically significant by T-test~\cite{David1997} at $p$-value$<0.05$.

\subsubsection{Accuracy of Duplicate Action Detection}
\label{sec:dedpu}
As discussed in Section~\ref{sec:enhancingtaskkg}, we introduce a new duplicate-action relationship in the knowledge graph, and infer duplicate actions in code comments and tutorial text by a BERT-based duplicate sentence classifier.
We label 968 pairs of comment-text duplication actions, with 484 positive and negative pairs respectively.
We perform 10-fold cross-validation to confirm the accuracy of duplicate action detection.
The duplicate sentence classifier achieves 0.79 precision, 0.93 recall and 0.85 F1.

\subsubsection{Accuracy of Action/Attribute/Relation Extraction}
We adopt the same statistical sampling method~\cite{Singh1996} used in~\cite{taskkg} to evaluate the accuracy of the extracted actions, attributes and relations.
Specifically, we sample and examine the minimum number MIN of data instances in order to ensure that the estimated accuracy is in 5\% error margin at 95\% confidence level.
MIN is determined by $n_0/(1+(n_0-1)/populationsize)$ where $n_0 = (Z^2*0.25)/e^2$, and $Z$ is the confidence level's z-score and $e$ is the error margin.

We extract 23,918 programming actions, including 2,959 from document headings, 19,425 from tutorial text and 1,534 from code comments.
Compared with the 11,519 leaf actions extracted from tutorial text~\cite{taskkg}, our knowledge graph contains 40\% more leaf actions due to the consideration of code comments and new forms of activity sentences.
Consider the total 135,928 sentences in tutorial text and 5,838 sentences in code comments, code comments are more action intensive than tutorial text.
We examine 376 actions extracted from tutorial text and 307 actions extracted from code comments.
The accuracy of actions from tutorial text is 88.5\%, with the accuracy 96.0\% and 93.1\% for the identified action verbs and objects.
Compared with~\cite{taskkg}, the accuracy degrades about 3\%.
This is because we consider non-imperative activity sentences which involve new action phrase patterns.
For actions from code comments, the accuracy is 95.8\%, with the accuracy 96.7\% and 98.7\% for the identified action verbs and objects.
The higher accuracy for actions from code comments than those from tutorial text is due to the imperative sentence structure often used in code comments. 

For each type of action attribute, we examine 384 instances.
For actions from tutorial text, the extraction accuracy for the API, location, condition, goal and code attributes are 87.5\%, 82.6\%, 96.0\%, 93.2\%, and 81.5\%, respectively.
For actions from code comments, we examine the API, location, condition and goal attributes, as linking a code fragment with actions from the preceding comment has no errors according to the common code commenting practice.
The accuracy for API, location, condition and goal are 81.2\%, 88.6\%, 97.3\%, and 95.0\% respectively.
These accuracy results are consistent with the accuracy results reported in~\cite{taskkg}.



For each type of relations, we examine 384 instances.
We achieve the accuracy 95.8\%, 86.4\%, 91.7\% for hierarchical, sibling and precede-follow relations respectively, which are consistent with the results in~\cite{taskkg}.
We also examine 149 instances of newly introduced duplicate-action relations.
The accuracy is 78.5\%.
The erroneous duplicate action relations are due to the false positive predictions by the duplicate comment-text action sentence classifier.

\noindent
\fbox{
	\begin{minipage}{8.2cm} \emph{Our open information extraction methods are accurate and can construct high-quality knowledge graph.}
\end{minipage}}

\subsection{RQ5: Effectiveness of Task Search}
\label{sec:effect}

In this RQ, we analyze if the recommended code examples by our approach contain code solutions for solving the programming questions on Stack Overflow, such as the examples in Figure~\ref{fig:tutorialexamples}.

\subsubsection{Experiment Setup}
\label{sec:rq2setup}

We consider all 20 questions investigated in our formative study (see Section~\ref{sec:formativestudy}).
We use the code snippet that the question asker posts in the question body as the code under development.
As our current tool analyzes only the code in a method, we test each method provided in the question body one at a time.
We assume two types of code input.
First, we use \textit{all code} in the method for matching code examples.
Second, we use some issue API(s) in the code that are mentioned in the question and/or best answer (e.g., requestLocationUpdates() in Q6) or appear in the provided stack trace (e.g., BitmapFactory.decodeFile() in Q3) for matching code examples.
This \textit{key-API} option simulates that the developer selects some key code elements for recommendation.
All-code option can be regarded as a special case of key-API option, in which all APIs in the code is selected.

We examine the top-3 code examples ranked by the code matching similarity function in Section~\ref{sec:codematching}.
If a code example contains the code solution in the best answer of the question, we consider this code example as an effective recommendation for solving the question.
We compute the top-3 accuracy (Acc) and precision (Pre).
Furthermore, we use the API(s) mentioned in the best answer to find all code examples in our knowledge graph using these APIs.
We examine these code examples to identify all code examples that contain code solution for the question.
This allows us to compute the recall (Rec) of top-3 recommended code examples.
We also compute F1 which is a harmonic mean of precision and recall.

\subsubsection{Code Matching Similarity Computation}

 \begin{table}
 	\small
 	\renewcommand\arraystretch{1.5}
 	\centering
 	\caption{Performance of Similarity Function}
 	\label{tab:exp2_1}
 	\setlength\tabcolsep{2pt}
 	\begin{tabular}{|l|llll|llll|}
 		\hline
 		\multirow{2}{*}{\textbf{Settings}}&\multicolumn{4}{c|}{\textbf{All-Code as Input}}&\multicolumn{4}{c|}{\textbf{Key-API as Input}}\\
 		&\textbf{Acc}&\textbf{Pre}&\textbf{Rec}&\textbf{F1}&\textbf{Acc}&\textbf{Pre}&\textbf{Rec}&\textbf{F1}\\
 		\hline
		A-B-U&0.55&0.400&0.250&0.308&\textbf{0.75}&\textbf{0.550}&\textbf{0.381}&\textbf{0.450}\\
		\hline
		A-S-U&0.55&0.383&0.250&0.303&0.75&0.533&0.381&0.445\\
		\hline
		A-B-M&0.55&0.350&0.244&0.288&0.70&0.450&0.363&0.402\\
		\hline
		A-S-M&0.55&0.383&0.245&0.299&0.75&0.500&0.367&0.424\\
 		\hline
 		C-B-U&0.55&0.383&0.250&0.303&0.75&0.517&0.337&0.436\\
 		\hline
 		C-S-U&\textbf{0.55}&\textbf{0.400}&\textbf{0.252}&\textbf{0.310}&0.75&0.533&0.380&0.444\\
 		\hline
 		C-B-M&0.55&0.367&0.244&0.293&0.75&0.483&0.367&0.417\\
 		\hline
 		C-S-M&0.55&0.400&0.250&0.308&0.75&0.533&0.377&0.442\\
 		\hline
 	\end{tabular}
 	\vspace{-5mm}
 \end{table}

As mentioned in Section~\ref{sec:codematching}, our code matching similarity function has three variation points: 
specific \underline{A}PIs versus declaring \underline{C}lasses; 
\underline{B}ag versus \underline{S}et of APIs; and
considering \underline{U}matched APIs or considering only \underline{M}atched APIs.
Therefore, we have 8 settings: A-B-U, A-S-U, A-B-M, A-S-M, C-B-U, C-S-U, C-B-M, C-S-M.

Table~\ref{tab:exp2_1} presents the average top-3 accuracy, precision, recall and F1 over the 20 questions under these 8 settings.
For all-code as input, the top-3 accuracy is 0.55 for all similarity function settings.
That is, for 11 out of the 20 questions, our approach can recommend at least one code example containing effective code solution in the top-3 recommended code examples.
For key-API as input, the top-3 accuracy is 0.75 for all similarity function settings (i.e., at least one effective recommendation in the top-3 recommended code examples for 15 out of the 20 questions)
Different code matching settings affect the rankings of specific code examples, which result in certain variations in precision, recall and F1.
Considering the high top-3 accuracy, these ranking variations would not significantly affect the developer's use of the recommended code examples.
Overall, considering unmatched APIs is beneficial than considering only matched APIs.
Matching by bag of APIs (i.e., considering API usage times) is not beneficial than matching by set of APIs.
When the other two settings are the same, matching by declaring classes is in general on-par with or better than matching by specific APIs.

\subsubsection{Detailed Analysis of Recommended Code Examples}
\label{sec:recommendationanalysis}

The column CueMeIn Recommendation in Table~\ref{tab:formativestudy-rq3} summarizes the recommendation results by our approach for the 20 questions under the code matching setting A-B-U.
For each question, it lists the programming action in the first effective recommendation.
Rank-x is the rank of this recommendation.
Input=AllCode or KeyAPI means that this programming action is recommended when all code or some APIs in the question are used for code matching.
The symbol ``-'' indicates no effective recommendation. 

Our approach recommends programming actions with effective code solutions for all six questions (Q1/Q2/Q4/Q6/Q14/Q15) that activity-centric search by natural language queries finds solution-specific actions (SSAction).
In addition, our approach recommends effective code solutions for nine more questions (Q3/Q5/Q9/10/Q13/ Q16/Q17/Q19/Q20).
Among the 15 questions that our approach recommends effective code solutions, 
our approach is effective for 11 questions when simply taking the code snippets in the questions for code matching (i.e., Input=AllCode).
For the other four questions, our approach is effective when using specific APIs mentioned in the questions for code matching (i.e., Input=KeyAPI). 
For 9 out of the 15 questions, the top-1 recommendation contains effective code solution.
For 5 questions, the top-2 recommendation is effective, and for 1 question, the top-3 recommendation is effective.

Our approach fails to recommend effective code solutions for 5 questions (Q7/Q8/Q11/Q12/Q18).
As explained in Section~\ref{sec:formativestudy-rq1}, Q11 asks for an alternative API for an deprecated API, which is not usually in programming tutorials.
For Q7, Android Developer Guides describe only a solution by configuring the XML manifest file.
However, our tool currently considers only non-XML code solutions.
For Q8, the code snippet in the question is too short for effective code matching.
For Q12, the question asker uses BitmapFactory.decodeStream(), while the code solution in the tutorial uses BitmapFactory.decodeFile().
Although the two APIs support similar functionality, our tool currently does not analyze such similar APIs for code matching.
For Q18, the question asker provides the onCreate() method, but not the key API onSaveInstanceState(). 
As onCreate() is very common in the tutorials, our approach cannot find effective code solution related to the use of onSaveInstanceState().

To understand the capability of our approach, we compare the programming actions recommended by our approach and the tutorial links referenced in the best answers.
The results are shown in TaskRel.
For one of the 15 questions (Q16) that our approach makes effective recommendations, the recommended programming action is the same as (TaskRel=Same) the tutorial link recommended in the best answer.
For four of these 15 questions (Q9/Q13/Q19/Q20), the recommended programming actions are more fine-grained (TaskRel=Subtask) than the tutorial link recommended in the best answer.
That is, developers would suffer from less information overload with our recommended programming actions.
For three of the 15 questions (Q3/Q4/Q15), our approach recommends programming actions related to (TaskRel=Related) those recommended in the best answers, for example ``Decode a scaled Bitmaps'' recommended by our approach versus ``Loading Large Bitmaps'' in the best answers for Q3/Q4.
Note that both actions contain effective code solution to Q3/Q4.
For the rest seven questions (Q1/Q2/Q5/Q6/Q14/Q17), our approach recommends programming actions that have no relation (TaskRel=No) with the tutorial link recommended in the best answers.
However, our recommended programming actions also contain effective code solutions.

Finally, we analyze the role of comment-level actions in the recommendations.
The results are shown in CommentAns.
For four questions (Q3/Q6/Q16/Q17) (CommentAns=Rank-2 or Rank-3), our approach recommends comment-level actions (three at rank-2 and one at rank-3) whose associated code snippets contain effect code solutions.
Furthermore, for four other questions (Q4/Q9/Q14/Q19) (CommentAns=HL), although relevant comment-level actions are not recommended in the top 3, they are contained in the tutorial excerpt of the recommended programming actions, such as the examples in Figure~\ref{fig:tutorialexample2} and Figure~\ref{fig:tutorialexample4}.
Our tool highlights relevant comment-level actions (e.g., ``remove dialog title'' in Figure~\ref{fig:ui}) when the developer is inspecting the recommended programming action.
Such highlights help to reduce code information overload.

\noindent
\fbox{
	\begin{minipage}{8.2cm} \emph{Our code matching based task search can effectively recommend code solutions to the top-viewed programming questions on Stack Overflow, comparable to the code solutions provided in the best answers. Our task search method provides more specific solutions than Google search, and are much more effective than activity-centric search by natural language queries. Enriching the knowledge graph with comment-level actions and code snippets helps to improve action coverage and reduce code information overload.}
\end{minipage}}

\subsection{RQ3: Usefulness of CueMeIn Tool}

We conduct a small-scale user study to investigate the usefulness of our CueMeIn tool, compared with search-based methods.

\begin{table}
	\small
	\centering
	\caption{Six Android Bug-Fixing Tasks in User Study}
	\label{tab:searchtasks}
	\begin{tabular}{|l|l|}
		\hline
		
		1.How to get String resource in Android? \textit{Set text of} & easy\\
		\textit{\hspace{2mm} TextView to pre-defined String resource}&\\
		\hline
		
		2.How to display Toast in Android? \textit{Set duration to} & easy\\
		\hspace{2mm}\textit{LENGTH\_LONG, gravity to center}&\\
		\textit{\hspace{2mm}and text to ``I love this plugin!''}&\\
		\hline
		
		3.How to hide the title space for the Dialog & medium\\
		\hline
		
		4.How to solve \textit{IndexOutOfBoundEeception} InputStr- & medium\\
		\hspace{2mm}eam for Android? \textit{Assume you want to read the file}&\\
		\hspace{2mm}\textit{and store texts into buffer}&\\
		\hline

		5.How to load a very large image to Andorid? & hard\\
		\hspace{2mm}\textit{There is ``Allocation failed for scaled bitmap''}&\\ 
		\hspace{2mm} \textit{error when loading image directly}&\\ 
		\hline
		
		6.How to load information from Preference Activity? &hard\\
		\hspace{2mm}\textit{Assume you have a preference loading by }&\\
		\hspace{2mm}\textit{PreferenceActivity, and you want to load}&\\
		\hspace{2mm}\textit{the preference from another activity}&\\
		\hline
	\end{tabular}
	\vspace{-6mm}
\end{table}

\subsubsection{User Study Design}


Table~\ref{tab:searchtasks} lists the six Android app bug fixing tasks in our study.
Task 1/2/3/5/6 are designed from the Q5/Q1/Q9/Q3/Q6 in Table~\ref{tab:formativestudy} respectively.
We select these five questions because they cover different properties of our tool's recommendations: the effective input type, the rank positions of the program actions containing effective code solutions, and the role of comment-level actions.
Furthermore, these five questions have high view counts and different task difficulties.
Q3 and Q5 are also representative of other programming issues with similar underlying causes and solutions (i.e., Q3/Q4/Q12, Q5/Q15).
Task 4 is designed from the question \href{https://stackoverflow.com/questions/9171714/indexoutofboundsexception-when-read-and-write-from-standard-i-o}{\textcolor{blue}{``IndexOutOfBoundsException When read and write from Standard I/O''}} not included in our formative study.
The IndexOutOfBoundException in this question is caused by the erroneous parameter usage of InputStream.write().
As our current tool does not provide any specific support to API parameter misuse, Task 4 allows us to see the capability boundary of our tool.
We use the code snippets in the questions as buggy methods to fix and compose task requirements from the question descriptions.

We recruit 12 master students from our school who are familiar with Java development but have only beginner level of knowledge for Android development.
They simulate the developers who often encounter the issues like those in our formative study.
We confirm that none of the participants know the solutions to the buggy code in our experimental tasks. 
Based one the pre-experiment survey of the participants' programming experience, we randomly assign them into two comparable groups.
The experimental group uses the recommendations by our CueMeIn tool to complete the tasks, while the control group uses search engines (e.g., Google) the participants prefer to find solutions.
We require the control group not to view the original Stack Overflow questions from which our tasks are designed, but they can view any other web pages as they want. 

Before the experiment, we give the experimental group a 15-minutes tutorial about our tool's UI and usage.
By an easy pilot task, we ensure that the participants know how to interact with and interpret the CueMeIn's recommendations.
As the control group uses their familiar search engine, no training is needed.
The participants are given 15 minutes for each task.
They can modify the code as they wish to accomplish the task.
If the participants believe they complete the task, they can submit the code solutions before 15 minutes.
We record the task completion time.
A task is considered as a success if the submitted solution matches the solution provided in the best answers of the Stack Overflow questions.
If the participants cannot complete the task within 15 minutes, the current task is considered as a failure and they move to the next task.
At the end of each task, the participants are asked to rate the task difficulties.
The experiment group also rate the usefulness of our tool's recommendations.
The ratings use 5-point likert scale, with 1 being the lowest and 5 being the highest.
The participants use screen recorders to record the task completion process to assist the observation of their programming behaviors.

\subsubsection{Results and Findings}

\begin{table}
	\small
	\centering
	\caption{Performance Comparison}
	\vspace{0mm}
	\label{tab:performance}
	\begin{tabular}{|c|c|c|c|c|}
	\hline
	
	\multirow{3}{*}{Index}&\multicolumn{2}{c|}{Experimental Group}&\multicolumn{2}{c|}{Control Group}\\
	\cline{2-5}
	&AveTime (s)&\# Success&AveTime (s)&\# Success\\
	\hline
	P1&571.3&5&585.7&3\\
	\hline
	P2&474.2&5&809.3&2\\
	\hline
	P3&440.0&6&578.3&4\\
	\hline
	P4&492.3&4&833.8&1\\
	\hline
	P5&511.3&4&671.5&3\\
	\hline
	P6&536.7&6&682.3&5\\
	\hline
	Ave$\pm$&\multirow{2}{*}{504.3$\pm$46.4}&&\multirow{2}{*}{693.5$\pm$108.3}&\\
	stddev&&&&\\
	\hline

	\end{tabular}
	\vspace{-6mm}
\end{table}

Table~\ref{tab:performance} shows the average task completion time (AveTime) in seconds and the number of successful tasks (\#Success) per participant.
Overall, the participants in the experimental group complete the tasks faster than those in the control group ($504.3 \pm 46.4$ seconds versus $693.5 \pm 108.3$ seconds).
On average, the experimental group uses 14\%-60\% less time than the control group on each task.
All participants in the experimental group complete at least four tasks.
Two of them complete all six tasks.
In contrast, only two participants in the control group complete four or five tasks, and two participants complete only one or two tasks.
Task-1 is an easy task that all participants successfully complete, but the experimental group on average uses 60\% less time.
For Task-2 and Task-3, the two groups have comparable success rates (six in the experimental group versus 4 or 5 in the control group), but the experimental group completes the tasks 33\% and 44\% faster.
Only one participant in the experimental group fails on Task-5 and one fails on Task-6.
In contrast, four in the control group fails on Task-5 and none of the six participants complete the Task-6 in 15 minutes.
Furthermore, the experimental group uses 22\% and 38\% less time on Task-5 and Task-6 respectively, compared with the control group.
Neither group performs well on Task-4 (four failures in the experimental group versus five failures in the control group), and they have the least (14\%) completion time difference. 

As the experimental tasks involve programming issues that developers frequently encounter, there are many resources on the Web (technical blogs, similar questions on Stack Overflow, or Q\&As on other forums) that contain the effective solutions.
However, observing the task completion videos reveals two difficulties faced by the control group in finding and using these resources effectively.
First, the participants have to formulate good natural language queries describing the task needs and program bugs.
Second, given a relevant page returned by the search engine, the participants have to read the page, digest its content, and finds the most relevant code buried in much irrelevant information with little hints which may be relevant or irrelevant.
In contrast, the experiment group participant receive recommendations directly from the code they are working on, without the need to formulate queries.
Furthermore, our tool enriches the tutorial excerpt with visual cues based on the programming action information in the knowledge graph, which helps to judge the relevance of the recommended programming actions and code examples and spot the code solutions if any.

Out of the 36 ratings (6 tasks by 6 participants) of the usefulness of our tool's recommendations, 19 are rated as useful.
12 non-useful ratings are given for our tool's recommendations on Task-4 and Task-6.
This is because our tool's recommendations do not cover all necessary knowledge for the two tasks. 
The root cause of the bug in Task-4 is a parameter misuse.
Our tool actually recommends a code example with the correct parameter usage.
However, due to the lack of knowledge about the API definition, the participants do not notice the difference in parameter usage between the buggy code and the recommended code example.
For Task-6, our tool also recommends a relevant code example.
But Task-6 demands much knowledge about Android preference which is not present in the recommended programing task and code example.

Overall, we do not observe significant difference between the ratings of task difficulties by the two groups.
Our interviews with the participants suggest that this is because the participants rate the task difficulties mainly based on the knowledge they need to learn for the tasks.
Even though our tool may provide relevant and useful code examples, the process of learning necessary knowledge for the beginner-level developers remain largely the same.

\vspace{1mm}
\noindent\fbox{\begin{minipage}{8.4cm} \emph{The participants, prompted with the task-oriented API usage examples, complete the programming tasks faster and more successfully than those using search engine to search solutions.
The usefulness of our tool could be enhanced by incorporating other types of API knowledge in the knowledge graph and recommendation.} \end{minipage}}

\begin{figure}
	\centering
	\subfigure[Task Difficulty]{\includegraphics[width=.2\textwidth]{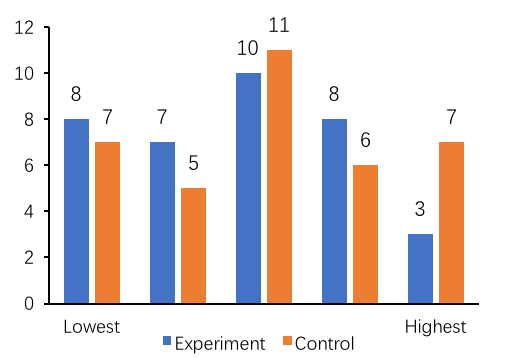}
		\label{fig:diff}	
	}
	\hfil
	\subfigure[Tool Usefulness]{\includegraphics[width=.2\textwidth]{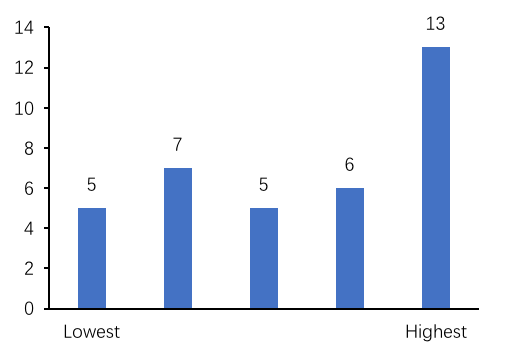}
		\label{fig:useful}	
	} 
	\vspace{-1mm}
	\caption{5 Likert Scale of Difficulty and Usefulness Marks}
	\vspace{-7mm}
	\label{fig:useful_diff}
\end{figure}


\section{Related Works}
Recently, Li et al.~\cite{icsme18} and Liu et al.~\cite{mingweiliu-fse2019} develop open information extraction methods to construct API knowledge graph from API reference documentation.
In these knowledge graphs, entities are APIs, edges are declaration relations between APIs, and entity attributes are API visibility and caveat sentences.
Our knowledge graph is is constructed from programming tutorials, in which entities are programming actions, edges are three types of action relationships, and entity attributes are action-related information and code.
The two types of knowledge graphs can be integrated through the APIs involved in programming actions.
This integrated knowledge graph could offer better knowledge support in programming tasks than each independently, for example for the Q11 and Task-4/6 that our knowledge graph offers limited support.

Our work extends the programming task knowledge graph~\cite{taskkg} with more actions and finer-grained code snippets.
Existing works on knowledge graph~\cite{icsme18, mingweiliu-fse2019, taskkg} support only natural language search of API and task knowledge.
In contrast, our work is the first to integrate the knowledge graph in the IDEs, and support the recommendation of programming actions by matching the code under development and the code examples in the knowledge graph.
Our tool adopts observe-push paradigm to prompt developers with API usage examples.
This paradigm is also used in other tools like Prompter~\cite{LucaPonzanelli-MSR2014}.
Different from our tool, Prompter retrieves pertinent posts from Stack Overflow.
Furthermore, Prompter just presents relevant posts with a relevance score, but our underlying knowledge graph allows us to annotate the recommended tutorial excerpt and code examples with rich action and API visual cues.

Some methods have been proposed to infer specifications from text, such as domain glossary~\cite{congwang-fse2019}, call-order constraints~\cite{TanSosp07, ZhongAse09}, parameter constraints~\cite{zhouyuicse2017apidirective}.
Different from these methods, our work constructs a knowledge graph of programming actions from programming tutorials to support task-oriented code search.
TaskNav~\cite{TaskNav15} also extracts task phrases from API tutorials to support tutorial paragraph retrieval by task phrase queries.
As our formative study shows, searching programming actions by natural language queries suffer from task mismatch and code information overload problems, which our tool aims to solve.

\vspace{2mm}
\section{Conclusion and Future Work}

Through a study of 20 top-viewed programming questions on Stack Overflow, this paper reveals two barriers (task mismatch and code information overload) to effective use of API usage examples in programming tutorials in programming tasks.
We show that existing document search and activity-centric search methods cannot effectively address these two barriers.
We present a knowledge graph based approach to add the barriers.
Our knowledge graph extends existing work with more actions, especially fine-grained actions and code snippets from code comments.
We adopt a completely different task search method based on code matching, and develop an IDE plugin to prompt developers with task-oriented API usage examples enriched with action information in the knowledge graph.
Our evaluation confirm the effectiveness and usefulness of our approach for finding code solutions in tutorials to programming questions developers often encounter.
To generalize our results and findings, we will apply our approach to more programming tutorials and release our knowledge graph and tool for public evaluation. 

\bibliographystyle{ACM-Reference-Format}
\bibliography{ase2020citelist.bib}

\end{document}